\documentclass[conference]{IEEEtran}
\IEEEoverridecommandlockouts
\usepackage{amsmath,amssymb,amsfonts}
\usepackage{algorithmic}
\usepackage{array}
\usepackage{hyperref}
\usepackage{booktabs}
\newcolumntype{C}[1]{>{\centering\let\newline\\\arraybackslash\hspace{0pt}}m{#1}}
\usepackage{cleveref}
\crefname{figure}{Fig.}{Figs.}
\Crefname{figure}{Fig.}{Figs.}
\crefname{table}{Table}{Tables}
\Crefname{table}{Table}{Tables}
\crefname{equation}{Eq.}{Eqs.}
\crefname{equation}{Eq.}{Eqs.}
\usepackage{graphicx}
\usepackage{multirow}
\usepackage{subcaption} 
\usepackage{textcomp}
\usepackage{xcolor}
\usepackage[normalem]{ulem}
\usepackage[style=ieee, maxnames=5, minnames=1]{biblatex}
\AtBeginBibliography{\footnotesize}
\def\BibTeX{{\rm B\kern-.05em{\sc i\kern-.025em b}\kern-.08em
    T\kern-.1667em\lower.7ex\hbox{E}\kern-.125emX}}

\begin{document}

\title{Evaluating the Influence of Temporal Context on Automatic Mouse Sleep Staging through the Application of Human Models\\
\thanks{J.G.C. is supported by the Danish Data Science Academy, which is funded by the Novo Nordisk Foundation (NNF21SA0069429) and VILLUM FONDEN (40516). A.N.Z. is funded by the Lundbeck Foundation (R347-2020-2439).

© 2024 IEEE. Personal use of this material is permitted. Permission from IEEE must be obtained for all other uses, in any current or future media, including reprinting/republishing this material for advertising or promotional purposes, creating new collective works, for resale or redistribution to servers or lists, or reuse of any copyrighted component of this work in other works.
}
}

\author{Javier García Ciudad\IEEEauthorrefmark{1}, Morten Mørup\IEEEauthorrefmark{2}, Birgitte Rahbek Kornum\IEEEauthorrefmark{1} and Alexander Neergaard Zahid\IEEEauthorrefmark{2} \\
\normalsize\IEEEauthorrefmark{1}Department of Neuroscience, University of Copenhagen, Copenhagen, Denmark \\
\IEEEauthorrefmark{2}Department of Applied Mathematics and Computer Science, Technical University of Denmark, Kgs. Lyngby, Denmark \\
\{jgciudad, kornum\}@sund.ku.dk \{mmor, aneol\}@dtu.dk
}

\maketitle

\begin{abstract}
In human sleep staging models, augmenting the temporal context of the input to the range of tens of minutes has recently demonstrated performance improvement. In contrast, the temporal context of mouse sleep staging models is typically in the order of tens of seconds. 
While long-term time patterns are less clear in mouse sleep, increasing the temporal context further than that of the current mouse sleep staging models might still result in a performance increase, given that the current methods only model very short term patterns.
In this study, we examine the influence of increasing the temporal context in mouse sleep staging up to 15 minutes in three mouse cohorts using two recent and high-performing human sleep staging models that account for long-term dependencies. These are compared to two prominent mouse sleep staging models that use a local context of 12 s and 20 s, respectively. 
An increase in context up to 28 s is observed to have a positive impact on sleep stage classification performance, especially in REM sleep. However, the impact is limited for longer context windows.
One of the human sleep scoring models, L-SeqSleepNet, outperforms both mouse models in all cohorts. This suggests that mouse sleep staging can benefit from more temporal context than currently used.

\end{abstract}


\section{Introduction}
Mouse models are commonly used in sleep research because of the availability of gene manipulation techniques to mimic human sleep disorders. Sleep staging is one of the main tools used to analyse sleep. In this process, electrophysiological recordings are divided into short windows (called sleep epochs) that are usually chosen to be between 4 s and 30 s in mice. The sleep epochs are then classified into one of three sleep stages: wake, rapid eye movement sleep (REM) and non-REM sleep (NREM). 

While deep learning models have been proposed for mice, research on automatic sleep staging has predominantly focused on humans. Consequently, human sleep staging benefits from a faster integration of the latest developments in deep learning. Despite differences between human and mouse sleep, such as the number of stages (humans exhibit three types of NREM, whereas mice have only one), both species share electrophysiological signatures during sleep \cite{bastianini2015recent}. Therefore, it is reasonable to look for inspiration from models developed for humans to enhance the automatic staging of mouse sleep.

Two of the latest advances that have led to state-of-the-art (SOTA) results in human sleep staging are the sequence-to-sequence framework and the transformer model \cite{phan2022automatic}. One of the reasons why these two methods have led to increased performance is that they are more suitable for modelling the long-term time dependencies present in human sleep, whereas other previously used architectures such as recurrent neural networks have a limited memory span \cite{liu2022attention}. Therefore, to predict the stage of a single sleep epoch, the SOTA human models incorporate information from tens of adjacent sleep epochs.

In contrast, the automatic mouse sleep staging models proposed to date include a few neighboring epochs at best, meaning that only short-term dependencies are taken into account. One reason for this is the use of simpler architectures in mouse sleep staging. A second reason is that it is not clear if long-term patterns exist in mice \cite{bastianini2015recent}. Mice have a polyphasic distribution of sleep (i.e., their sleep is distributed in multiple intervals throughout the 24-hour day) and their NREM-REM cycle is much shorter than humans' (around 10 minutes) \cite{rayan2022sleep}. However, considering that current methods only include the very local context, automatic mouse sleep staging might still benefit from including more temporal context.

In \citeauthor*{miladinovic2019spindle}~\cite{miladinovic2019spindle}, they showed how including a first-order hidden Markov model (HMM) to model one-step time dependencies improves the performance (especially in REM), which we corroborated in our experiments. However, increasing the order of the HMM to cover a long time range is computationally unfeasible. Instead, we explore the influence of longer temporal context in mice using models accounting for long-term dependencies developed for humans.

SleepTransformer \cite{phan2022sleeptransformer} is a transformer-based human sleep staging model, and L-SeqSleepNet \cite{phan2023seqsleepnet} is a subsequent model proposed by the same authors to increase the temporal context. We trained both models from scratch on mouse data and compared them to two mouse sleep staging models, SPINDLE \cite{miladinovic2019spindle} and \citeauthor*{grieger2021automated}~\cite{grieger2021automated}. Specifically, we investigate:
\begin{itemize}
    \item What impact do the sequence-to-sequence framework and the transformer architecture have on the performance when compared to SOTA mouse sleep staging models?
    \item What effect on performance does the inclusion of longer temporal contexts have on mouse sleep staging?
\end{itemize}

\section{Data}
For training, we use in-house data from the Kornum lab (University of Copenhagen), consisting of data from 33 healthy mice labelled by 5 sleep experts, with each sleep epoch having only one label from one of the experts. 
For testing, we used 4 mice from the Kornum lab, as well as 4 healthy mice from the Brown lab (University of Zürich), and 6 healthy mice from the Tidis lab (University of Bern). The data from the latter two were made available together with SPINDLE \cite{miladinovic2019spindle}, and they are double-scored by a pair of sleep experts from each lab~\cite{miladinovic2019spindle}.
All cohorts contain 2 EEG and 1 EMG channels, and the length of the sleep epochs is 4 s.
For an overview of the distribution of sleep stages in each dataset, see~\cref{tab:data_stats}.

\section{Methods}
\subsection{Models}
In this section, we briefly describe the models used for completeness. For an overview of the major differences, please see~\cref{tab:models}. However, we refer to the original publications for more specific details. The code is available at \url{https://github.com/jgciudad/TemporalContextEMBC24}.

\subsubsection{SleepTransformer~\cite{phan2022sleeptransformer}}
SleepTransformer was the first model to use the transformer architecture for human sleep staging, achieving SOTA performance. It is a sequence-to-sequence model that takes a sequence of $\mathit{L}$ sleep epochs as input, and predicts a sleep stage for each epoch in the sequence. The input to the model are spectrograms. The model consists of two main levels: the epoch transformer and the sequence transformer. In the epoch transformer level, the temporal dependencies inside each of the input sleep epochs are learned using a transformer encoder. The 2D output of the transformer encoder is then reduced to a vector representation by a weighted combination of the columns, where the weights are learned by an additive attention layer. The vector representations of the $\mathit{L}$ epochs in the input sequence are stacked together and processed by the sequence transformer, which models the temporal dependencies across the whole input sequence.

When the authors of SleepTransformer increased the input sequence length $\mathit{L}$ in SleepTransformer and SeqSleepNet \cite{phan2019seqsleepnet} to cover a whole human sleep cycle (around 90 minutes), they obtained no improvement and even deteriorated performance \cite{phan2023seqsleepnet}. They believed that the cause of this effect was model deficiency, since the existence of human sleep cycles is well-known. Therefore, they proposed L-SeqSleepNet \cite{phan2023seqsleepnet}, a model more suitable for long sequences with slightly better performance than SleepTransformer, SeqSleepNet, and other SOTA models. 

\subsubsection{L-SeqSleepNet~\cite{phan2023seqsleepnet}}
As in SleepTransformer, L-SeqSleepNet takes as input a sequence of $\mathit{L}$ sleep epochs in the form of spectrograms, and outputs a sequence of $\mathit{L}$ sleep stages. It also consists of two different levels to model intra- and inter-epoch dependencies. The intra-epoch dependencies are learned by a bidirectional long short-term memory unit (BLSTM). After that, the outputs are reduced with an additive attention layer as in SleepTransformer. In the inter-epoch level, the input sequence of length $\mathit{L}$ is broken down into $\mathit{B}$ non-overlapping subsequences of size $\mathit{K}$, where $\mathit{L} = \mathit{B} \cdot \mathit{K}$. Intra-subsequence and inter-subsequence modelling is then performed by a BLSTM unit.

\subsubsection{SPINDLE~\cite{miladinovic2019spindle}}
SPINDLE  was the first deep learning model for sleep staging in mice. The input to the model is the spectrogram of the signals. To include contextual information, the two neighboring epochs on each side of the epoch being predicted are appended, resulting in an input sequence of five sleep epochs (20 s).  The model consists of 3 main blocks: a convolutional neural network (CNN) that predicts the stage of the epoch at the center of the input; a hidden Markov model (HMM) that corrects the sequence of stages predicted; and a second CNN that predicts whether the input sleep epoch is an artifact. Since we do not address artifact detection, only the sleep stage prediction CNN and the HMM are considered in our work. In the HMM, expert knowledge is incorporated into the model by suppressing the transitions REM → NREM and WAKE → REM, since these are not feasible in healthy mice. This is achieved by zeroing-out the corresponding entries in the transition matrix of the HMM.

\subsubsection{\Citeauthor*{grieger2021automated}~\cite{grieger2021automated}}
The model proposed by \citeauthor*{grieger2021automated}~\cite{grieger2021automated} is a SOTA model proposed for mouse sleep staging. The input to the model is the raw signal from the sleep epoch being predicted, together with the two neighboring epochs from either side. It is a CNN-based model consisting of 8 convolutional layers followed by a classifier with 3 fully connected layers.

\begin{table}[tb]
    \centering
    \caption{Percentage of sleep epochs per stage and data cohort. Epochs labelled as artifacts are excluded. In Brown and Tidis cohorts, both scorers are shown separated by a slash.}

    \label{tab:data_stats}
    \setlength{\tabcolsep}{3pt}
    \begin{tabular}{@{}lcccccc@{}}
         \toprule
         \textbf{Lab} & \textbf{Mice} & \textbf{Scorers}  & \textbf{\# epochs} & \textbf{W (\%)} & \textbf{N (\%)} & \textbf{R (\%)} \\
         \midrule 
         \textbf{Kornum} & 41 & 5 & 745,256 & 54.8 & 38.9 & 6.3 \\
         \textbf{Brown} & 4 & 2 & 71,164/75,444 & 53.0/45.9 & 39.7/47.3 & 7.4/6.8 \\
         \textbf{Tidis} & 6 & 2 & 129,576/129,576 & 49.8/48.7 & 45.2/45.2 & 5.0/6.0 \\
         \bottomrule
         \multicolumn{7}{l}{W: Wake; N: non-REM; R: rapid eye movement (REM).}
    \end{tabular}
\end{table}
\begin{table}[tb]
    \centering
    \caption{Models comparison.}
    \label{tab:models}
    \setlength{\tabcolsep}{3pt}
    \begin{tabular}{@{}lrrrr@{}}
    \toprule
    \textbf{Model} & \textbf{Domain} & \textbf{Network} & \textbf{Input size (s)} & \textbf{\# params.} \\
    \midrule
    \textbf{SleepTransformer \cite{phan2022sleeptransformer}} & Human & Transformer & 12-900 & 11,334,790 \\
    \textbf{L-SeqSleepNet \cite{phan2023seqsleepnet}} & Human & BLSTM & 12-880 & 663,014 \\
    \textbf{SPINDLE \cite{miladinovic2019spindle}} & Mouse & CNN & 20 & 14,656,403 \\
    \textbf{\citeauthor*{grieger2021automated} \cite{grieger2021automated}} & Mouse & CNN & 12 & 641,594 \\
    \bottomrule
    \multicolumn{5}{l}{BLSTM: bidirectional long short-term memory.} \\
    \multicolumn{5}{l}{CNN: convolutional neural network.}
    \end{tabular}
\end{table}
\subsection{Model modifications}
With the exception of the experimental settings mentioned in this subsection, we used the same architecture and training parameters as in the original publications of the models.

\subsubsection{Input channels}
The results reported by \citeauthor*{phan2022sleeptransformer} in SleepTransformer \cite{phan2022sleeptransformer} and L-SeqSleepNet \cite{phan2023seqsleepnet}, and in \citeauthor*{grieger2021automated} \cite{grieger2021automated} were obtained with only 1 EEG channel, although the three models can handle a multi-channel input. The originally reported results of SPINDLE were based on 2 EEG and 1 EMG channels. We study the four models using all three channels available in our cohorts.

\subsubsection{Variation of the input sequence length}
In SleepTransformer, they tested the values \{11, 21, 31, 41, 51\} for the length of the input sequence ($L$) in human sleep scoring, finding 21 to be best. In our work, we expanded this list to \{3, 7, 11, 21, 31, 41, 61, 81, 101, 157, 225\}. Given a sleep epoch duration of 4 s, this spans from an input sequence of 12 s to 900 s (15 minutes). 
In L-SeqSleepNet, there are two parameters that determine the length of the input sequence: $\mathit{B}$ (the number of subsequences) and $\mathit{K}$ (the length of the subsequence). 
To minimize the search space, we chose to fix the length of the subsequences ($\mathit{K}$) to the value found to perform best in humans \cite{phan2023seqsleepnet}, $\mathit{K} = 10$. Then, we studied the effect of the sequence length by testing the following values of the number of subsequences ($\mathit{B}$): \{1, 2, 3, 4, 6, 8, 10, 16, 22\}. In addition, we also tested two unique subsequences ($\mathit{K} = 1$) with a length of 3 and 7 sleep epochs. This means that the sequences tested range from 12 s to 880 s (14 min and 40 s).

\subsubsection{Data standardization}
We applied per-recording and per-channel standardization regardless of the standardization approach originally used in each model.
Through this, we want the comparison between models to reflect 
the difference caused by modelling choices, such as the network type and the length of the input, rather than other factors \cite{katsageorgiou2015sleep}. 

\subsubsection{Weighted loss}
The performance of deep learning models for sleep staging in mice is significantly affected by the varying occurrence of the sleep stages (\cref{tab:data_stats}). This particularly impairs the performance in the REM stage. In SPINDLE, they weighted the loss during training to counteract the unbalanced class ratios, where the loss of observations from a less frequent class is artificially increased, and the opposite for more frequent classes. Here, we do the same in all four models. The loss of a sample belonging to sleep stage $s$ is multiplied by a weight $w_s = 1/n_s$, where $n_s$ is the number of samples that belong to stage $s$ in that specific mini-batch.

\subsubsection{Metric for best model selection}
We used balanced accuracy on the validation set to select the best model during training, which corresponds to the average recall across the 3 sleep stages. Again, the imbalance between classes is the reason for this choice.

\section{Results}

\begin{table}[tb]
    \centering
    \caption{Influence of HMM in SPINDLE~\cite{miladinovic2019spindle}.}
    \label{tab:spindleHMM}
    \footnotesize
    \begin{tabular}{@{}lcccc@{}}
        \toprule
          & \multicolumn{2}{c}{\textbf{Accuracy}} & \multicolumn{2}{c}{\textbf{F1 REM}} \\ \cmidrule(lr){2-3} \cmidrule(lr){4-5}
         \textbf{Data cohort} & \textbf{No HMM} & \textbf{HMM} & \textbf{No HMM} & \textbf{HMM} \\
        \midrule
        \textbf{Kornum} & $95.4 \pm 0.0$ & $95.4 \pm 0.1$ & $85.2 \pm 0.0$ & $87.3 \pm 0.8$ \\
        \textbf{Brown} & $95.1 \pm 0.4$ & $95.3 \pm 0.1$ & $84.8 \pm 0.0$ & $89.6 \pm 2.3$ \\
        \textbf{Tidis} & $91.9 \pm 0.7$ & $92.9 \pm 0.4$ & $70.8 \pm 4.0$ & $79.7 \pm 3.0$ \\
        \bottomrule
        \multicolumn{5}{l}{HMM: hidden Markov model; REM: rapid eye movement.}
    \end{tabular}
\end{table}

\begin{figure*}[tb]
     \centering
     \begin{subfigure}[t]{0.49\textwidth}
        \centering
        \includegraphics[width=\columnwidth]{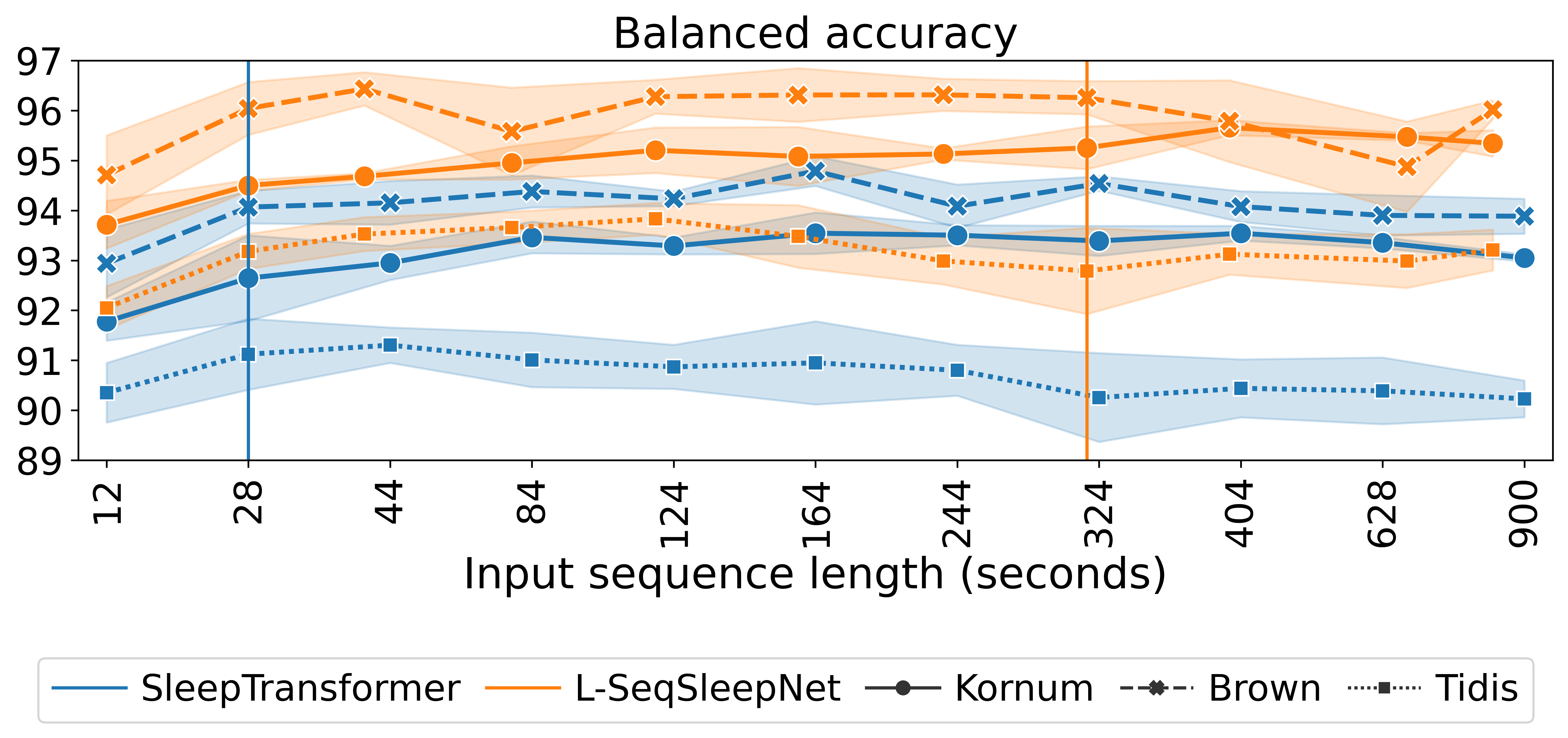}
        \caption{}
        \label{fig:bal_acc}
     \end{subfigure}
     \begin{subfigure}[t]{0.49\textwidth}
        \centering
        \includegraphics[width=\columnwidth]{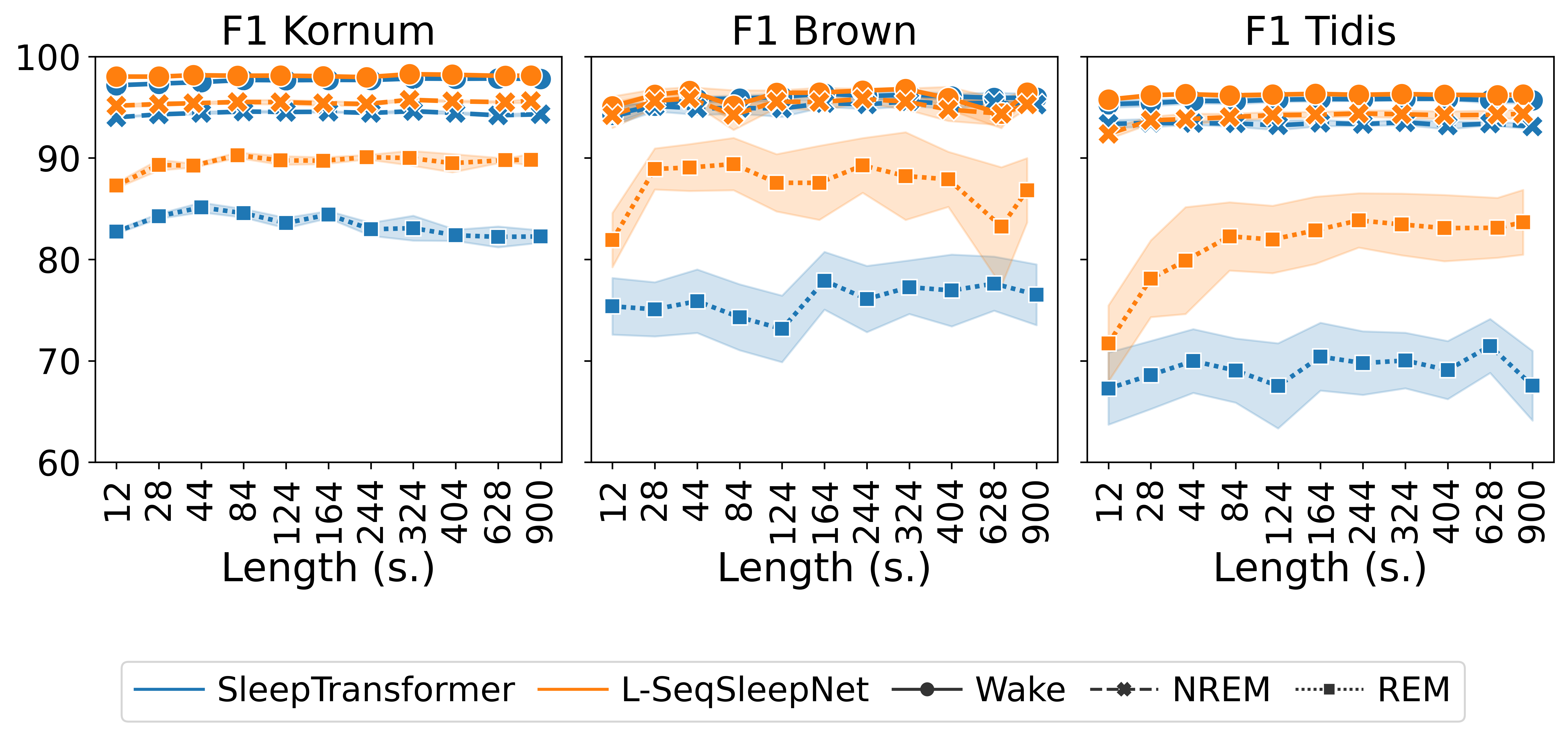}
        \caption{}
        \label{fig:f1}
     \end{subfigure}
        \caption{Balanced accuracy (a) and F1 score (b) of SleepTransformer and L-SeqSleepNet in test data across different input sequence lengths. The best sequence length on the validation set is highlighted with a vertical line. Average and standard deviation across iterations (and across scorers in the Brown and Tidis cohorts) is shown.}
        \label{fig:curves}
\end{figure*}

\begin{table*}[tb]
    \centering
    \caption{Performance of the models in test data. Average and standard deviation across iterations (and across scorers in the Brown and Tidis cohorts) is shown.}
    \label{tab:mouse_models}    
    \setlength{\tabcolsep}{2pt}
    \resizebox{\textwidth}{!}{
    \begin{tabular}{@{}l C{1.4cm} C{1.4cm} C{1.4cm} C{1.4cm} C{1.4cm} C{1.4cm} C{1.4cm} C{1.4cm} C{1.4cm} C{1.4cm} C{1.4cm} C{1.4cm}@{}}
        \toprule
        & \multicolumn{4}{c}{\textbf{Kornum}} & \multicolumn{4}{c}{\textbf{Brown}} & \multicolumn{4}{c}{\textbf{Tidis}} \\ \cmidrule(lr){2-5} \cmidrule(lr){6-9} \cmidrule(l){10-13}
        \textbf{Model} & \textbf{Bal. acc.} & \textbf{F1-W} & \textbf{F1-N} & \textbf{F1-R} & \textbf{Bal. acc.} & \textbf{F1-W} & \textbf{F1-N} & \textbf{F1-R} & \textbf{Bal. acc.} & \textbf{F1-W} & \textbf{F1-N} & \textbf{F1-R} \\
        \midrule
        \textbf{SleepTrans.~\cite{phan2022sleeptransformer}} & $92.6 \pm 0.9$ & $ 97.3 \pm 0.0 $ & $ 94.3 \pm 0.1$ & $ 84.3 \pm 0.2$ & $94.1 \pm 0.3$ & $95.2 \pm 0.3$ & $95.1 \pm 0.5$ & $75.1 \pm 2.7$ & $91.1 \pm 0.7$ & $95.4 \pm 0.3$ & $93.5 \pm 0.4$ & $68.6 \pm 3.4$ \\
        \textbf{L-SeqSleepNet~\cite{phan2023seqsleepnet}} & $\mathbf{95.3 \pm 0.4}$ & $\mathbf{98.3 \pm 0.1}$ & $\mathbf{95.7 \pm 0.1}$ & $\mathbf{90.0 \pm 0.7}$ & $\mathbf{96.3 \pm 0.3}$ & $\mathbf{96.8 \pm 0.1}$ & $\mathbf{95.6 \pm 0.8}$ & $88.2 \pm 4.3$ & $\mathbf{92.8 \pm 0.9}$ & $\mathbf{96.3 \pm 0.1}$ & $\mathbf{94.3 \pm 0.4}$ & $\mathbf{83.4 \pm 3.0}$ \\
        \textbf{SPINDLE~\cite{miladinovic2019spindle}} & $92.7 \pm 0.5$ & $97.1 \pm 0.1$ & $94.2 \pm 0.1$ & $87.3 \pm 0.8$ & $96.0 \pm 0.1$ & $96.4 \pm 0.1$ & $95.1 \pm 0.4$ & $\mathbf{89.6 \pm 2.3}$ & $91.8 \pm 0.4$ & $94.5 \pm 0.3$ & $93.1 \pm 0.4$ & $79.7 \pm 3.0$ \\
        \textbf{\citeauthor*{grieger2021automated}}~\cite{grieger2021automated} & $ 93.7 \pm 0.2$ & $97.7 \pm 0.1$ & $94.0 \pm  0.4$ & $82.5 \pm 2.6$ & $87.0 \pm 2.7$ & $94.2 \pm 0.1$ & $82.1 \pm 5.2$ & $53.1 \pm 9.2$ & $66.2 \pm 5.3$ & $43.2 \pm 14.2$ & $77.1 \pm 6.7$ & $37.0 \pm 7.5$ \\
        \bottomrule
        \multicolumn{13}{l}{W: wake; R: rapid eye movement sleep (REM); N: non-REM.}
    \end{tabular}
    }
\end{table*}

All models underwent three training iterations from scratch to account for parameter initialization and training stochasticity. The reported results show the average and standard deviation across these three iterations. In the Brown and Tidis cohorts, the performances against both sleep experts were also averaged. We initially confirm the improvement in the REM stage using a HMM, as initially reported in SPINDLE~\cite{miladinovic2019spindle} (see \cref{tab:spindleHMM}).

In~\cref{fig:curves}, we compare how SleepTransformer and L-SeqSleepNet perform with different sequence lengths. A small increase in temporal context from 12 s to 28 s improved the balanced accuracy in both models (\cref{fig:bal_acc}). However, longer temporal context does not seem to have an influence, as the curves remain mostly flat after this step. In \cref{fig:f1}, the sequence length has a much larger positive impact on REM stage classification, which was also observed when adding the HMM in SPINDLE~\cite{miladinovic2019spindle}. The improvement in REM is especially pronounced in the Brown and Tidis cohorts, with the latter showing further improvement when increasing the context to 44 s. This could mean that temporal context is even more important for predicting REM in challenging situations, such as when presented with data from an unseen cohort.

The sequence length with the best balanced accuracy on the validation set was selected for comparison with the two models from mice. These are the sequence lengths of 28 and 324 s for SleepTransformer and L-SeqSleepNet, respectively, which are highlighted with a vertical line (\cref{fig:bal_acc}). However, the performance in the validation set showed extremely low variation across the sequence length, especially in L-SeqSleepNet. This explains why the best sequence lengths in SleepTransformer and L-SeqSleepNet are so far apart. The larger standard deviation in the Brown and Tidis cohorts can be explained both by the increase in variability between training runs in new cohorts, and to the presence of two scorers. 

The performance of all four models is shown in~\cref{tab:mouse_models}. We observe that the current SOTA in mouse sleep staging is outperformed in almost all metrics by using a model with longer context, L-SeqSleepNet. However, longer context and a more advanced architecture does not necessarily lead to an improvement, as SPINDLE slightly outperforms SleepTransformer.
The performance of the model from \citeauthor*{grieger2021automated} decreases dramatically in the Brown and Tidis cohorts, indicating poor generalization to data cohorts different to the training cohort. Considering that SPINDLE, a relatively similar model in terms of architecture, shows much better generalization, it could indicate either a generalization disadvantage, or that more training is needed when using raw signals.

\section{Discussion}
The influence of the sequence length in automatic sleep staging has only been studied in some human models \cite{phan2022sleeptransformer, phan2023seqsleepnet, phan2019seqsleepnet, supratak2017deepsleepnet}. However, these studies only tested a limited number of values. In addition, in some cases the selection was based on a single training and the performance of all tested values was not reported. Although in mice, our results show a high standard deviation across multiple training runs, highlighting the importance of using multiple training runs or cross-validation in the hyperparameter selection.

The lower parameter count in L-SeqSleepNet than in SPINDLE may contribute to its better performance (see \cref{tab:models}). However, SPINDLE outperforms SleepTransformer while having more parameters, especially in the REM stage and in the Brown and Tidis cohorts. This could be due to transformer-based architectures requiring more data, which increases the likelihood of overfitting \cite{phan2023seqsleepnet}. SleepTransformer's need for more data could  explain why it does not outperform SPINDLE despite the longer context, and why it is less influenced by the change in temporal context than L-SeqSleepNet.

\section{Conclusion}

Sleep stage scoring in mice can benefit from sequence-to-sequence modeling as the L-SeqSleepNet yields superior performance over SOTA mouse sleep staging models. However, transformer-based models may need additional data in order to reach SOTA performance, which should be researched in future studies.
Increasing the temporal context up to 28 s positively impacts sleep scoring performance with limited impact above that limit.
These results indicate that future mouse models may positively benefit from looking at longer contexts rather than simply looking at the current or immediately neighboring epochs.

\printbibliography

\end{document}